\documentclass[aps,prl,twocolumn,superscriptaddress]{revtex4-1}

\usepackage{graphicx}
\usepackage{epstopdf}
\usepackage{amsmath}
\usepackage[usenames,dvipsnames]{color}
\usepackage{adjustbox}
\usepackage{tabularx}

\usepackage[normalem]{ulem}

\newcommand*{\ora}{\overrightarrow}

\begin{document}

\title{Relevance of Shear Transformations in the Relaxation of Supercooled Liquids}

\author{Matthias Lerbinger}
\author{Armand Barbot}
\author{Damien Vandembroucq}
\author{Sylvain Patinet}
\email[]{sylvain.patinet@espci.fr}
\affiliation{PMMH, CNRS, ESPCI Paris,  Universit\'e PSL, Sorbonne Universit\'e,
 Universit\'e Paris Cit\'e, 75005 Paris, France}

\date{\today}

\begin{abstract}
While deeply supercooled liquids exhibit divergent viscosity and increasingly heterogeneous dynamics as the temperature drops, their structure shows only seemingly marginal changes. Understanding the nature of relaxation processes in this dramatic slowdown is key for understanding the glass transition. Here, we show by atomistic simulations that the heterogeneous dynamics of glass-forming liquids strongly correlate with the local residual plastic strengths along soft directions computed in the initial inherent structures. The correlation increases with decreasing temperature and is maximum in the vicinity of the relaxation time. For the lowest temperature investigated, this maximum is comparable with the best values from the literature dealing with the structure-property relationship. However, the nonlinear probe of the local shear resistance in soft directions provides here a real-space picture of relaxation processes. Our detection method of thermal rearrangements allows us to investigate the first passage time statistics and to study the scaling between the activation energy barriers and the residual plastic strengths. These results shed new light on the nature of relaxations of glassy systems by emphasizing the analogy between the thermal relaxations in viscous liquids and the plastic shear transformation in amorphous solids.
\end{abstract}


\maketitle

When cooled fast enough, most substances avoid crystallization and form a glass~\cite{cavagna_supercooled_2009}. Before reaching the glass transition, they are trapped in a supercooled liquid metastable state, with a diverging relaxation time as the temperature is lowered, and show complex dynamics~\cite{keys_excitations_2011} marked by the
presence of dynamical heterogeneities~\cite{ediger_spatially_2000}. A fundamental problem in condensed matter that has remained unsolved so far is that it is extremely difficult to identify any structural signature associated with this dramatic change in dynamics. Still, a clear link between structure and dynamics is implied by the increased
sensitivity of the dynamics to the initial structural conditions as the temperature decreases~\cite{widmer-cooper_how_2004}. Recent major advances rely on the identification of the so-called locally favored
structures~\cite{malins_identification_2013} and indicate the necessity of employing a locally coarse-grained description of the structure~\cite{berthier_structure_2007,tong_revealing_2018,boattini_averaging_2021}. Remarkable progress has also been achieved using multibody order parameters~\cite{tong_structural_2019,tong_role_2020} or advanced machine learning methods~\cite{schoenholz_structural_2016,bapst_unveiling_2020,boattini_autonomously_2020,paret_assessing_2020,sharma_identifying_2022}.

Although these different approaches have established an unprecedented link between local dynamics and structure, most rely upon scalar quantities, i.e., they disregard the orientation dependencies. Furthermore, the underlying mechanisms of the relaxation processes have remained elusive. Contrasting results have been reported in the literature showing
that thermally activated excitations could take the form of single atomic jumps~\cite{vollmayr-lee_single_2004},
stringlike motions~\cite{donati_spatial_1999,vogel_particle_2004,ji_geometry_2022} or local shears~\cite{widmer-cooper_central_2009}. In all cases, the spatial structure of the relaxations tends to localize with decreasing temperature~\cite{coslovich_localization_2019,shimada_spatial_2021}.

A promising approach to understanding the relaxation of supercooled liquids consists of reversing the problem by considering highly viscous liquids as flowing solids~\cite{dyre_colloquium_2006}. The rationale behind this approach relies upon the description of liquids from a potential energy landscape point of view~\cite{goldstein_viscous_1969,debenedetti_supercooled_2001}. In this picture, for sufficiently low temperatures, the system spends most of its time vibrating around potential energy minima before
quickly jumping to other minima. A succession of Inherent States (ISs) can thus describe the dynamics~\cite{doliwa_what_2003}. More than a mere simplification, this solidlike description has enabled the understanding of key aspects of glass-forming liquids from a mechanical description of their ISs~\cite{heuer_exploring_2008,del_gado_nonaffine_2008}. Perturbative approaches based on soft vibrational
modes have been linked to irreversible rearrangements~\cite{widmer-cooper_irreversible_2008}, a non-Arrhenian
characteristic energy scale has been identified in fragile liquids~\cite{lerner_characteristic_2018,kapteijns_does_2021} and the presence of long-range anisotropic stress correlations explained~\cite{lemaitre_structural_2014}. All of these studies point to the importance of soft localized excitations. However, a direct
link between relaxations in liquids and the local nonlinear plastic rearrangements of ISs remains to be established~\cite{li_local_2022}.

This Letter aims to demonstrate how local soft directions in the initial ISs account for the local relaxation rates of the parent supercooled liquids. For this, we rely on a local shear test (LST) numerical method providing access to the local residual plastic strength $\Delta\tau^{c}$~\cite{patinet_connecting_2016}. This structural indicator has proved to be very helpful in athermal amorphous solids for capturing the barrier dependencies to preparation~\cite{barbot_local_2018}, shear banding~\cite{barbot_shearband_2020}, anisotropy~\cite{patinet_origin_2020} and to predict plastic activity~\cite{richard_predicting_2020}. 

We analyze molecular dynamics (MD) simulations of supercooled liquids equilibrated at different temperatures from which
temporal series of ISs are generated. A strong connection is established between the local residual plastic strength computed in the initial ISs along the weakest shear direction and the local dynamics. The maximum correlation is found around the system relaxation time $t_{\alpha}$ and increases as the temperature decreases. Additionally, a remarkable correlation is established between the core displacement fields of athermal plastic and thermal rearrangements, respectively observed in LST and MD. The thermal excitations are shown to be more frequent in the softest shear directions, thus providing a real-space picture of the relaxation processes. Local relaxation locations and first passage times are detected using the linear response of underlying ISs. This method allows us to investigate the scaling relation between the energy barriers and the residual plastic strengths~\cite{johnson_universal_2005,maloney_energy_2006,Egami-NatComm14}.

\begin{figure}
   \includegraphics[width=0.95\columnwidth]{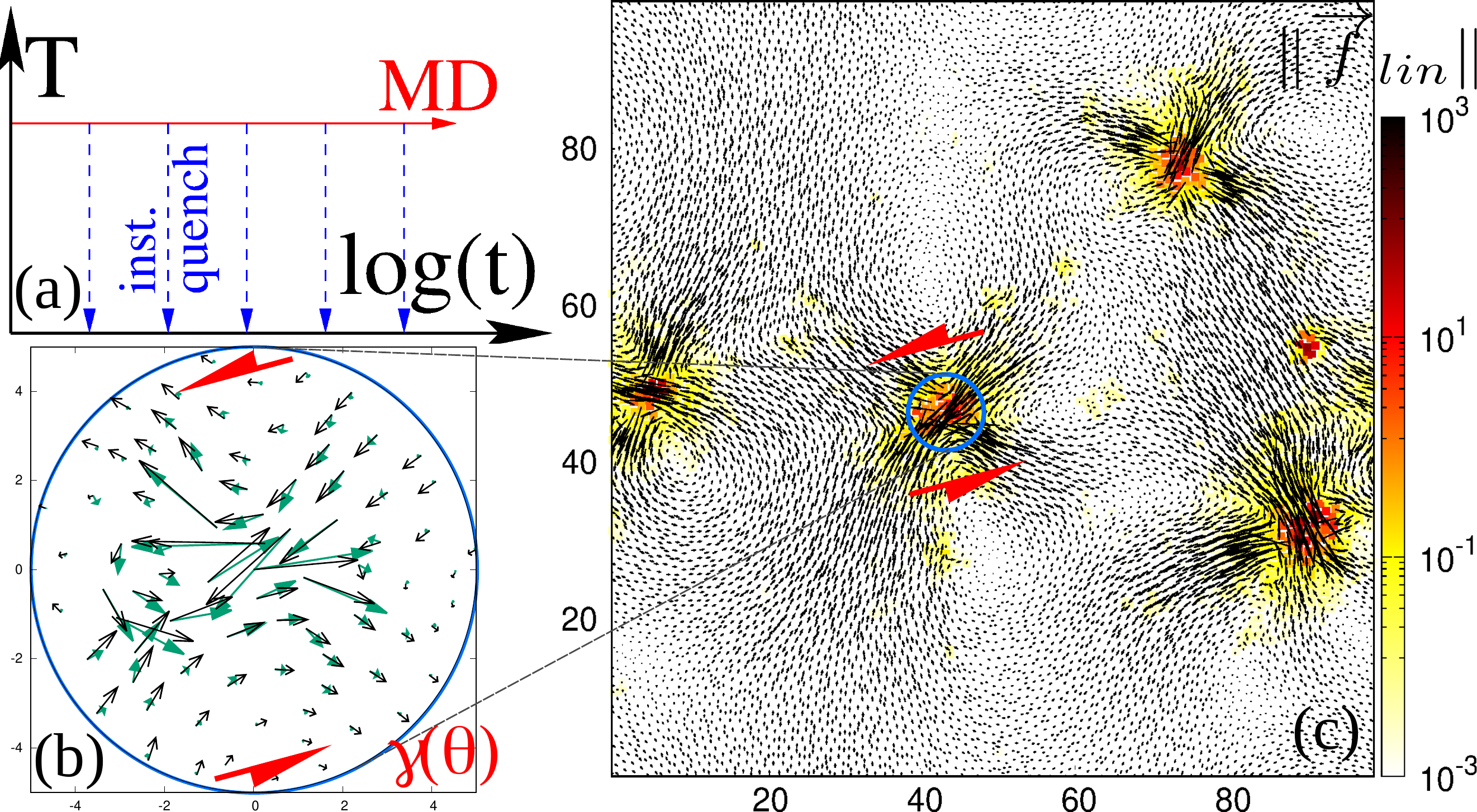}
\caption{\label{fig:method} Methods: (a) Generation of ISs by instantaneous quench from MD trajectories. $\Delta\tau^{c}(\protect\ora{r},\theta)$ are sampled in the initial ISs at $t=0$. (b) Normalized thermal rearrangement core displacement field $\protect\ora{u}_{th}$ (black arrows) between two ISs separated by $\Delta t=0.1$ from a liquid at $T=0.32$. The corresponding normalized LST displacement field $\protect\ora{u}_{LST}$ (cyan arrows) that maximizes the correlation over $\theta$ is found in the vicinity of the softest direction. (c) Corresponding far displacement (arrows) and linear response $\lVert \protect\ora{f}_{lin}\rVert$ (colors) fields.}
\end{figure}
 
{\em Simulation methods. - }The systems under study are two-dimensional binary mixtures~\cite{barbot_local_2018}. A sample contains $10^{4}$ atoms interacting via Lennard-Jones (LJ) potentials vanishing at a cutoff distance $R_{cut}=2.5$ (see Supplemental Material~\cite{SM}). MD simulations are performed in the NVT ensemble in square boxes equipped with periodic boundary conditions. In the following, all observables are expressed in LJ units.

The results that we analyze are taken from $20$ independent samples simulated at three temperatures $T=0.5$, $0.351$, and $0.32$, respectively located around the onset temperature, the mode-coupling crossover and in a regime below which the MD equilibrium sampling starts to be numerically untractable. The systems are equilibrated such that the potential energy is stationary over a time $100\,t_{\alpha}$. The relaxation time $t_{\alpha}$ is estimated from the self-intermediate scattering function $F_{s,CR}(q,t)$ using the first neighbor cage-relative (CR) displacements $\Delta \protect\ora{r}_{CR}$~\cite{shiba_unveiling_2016} such that $F_{s,CR}(q\!=\!2\pi/a,t\!=\!t_{\alpha})=1/e$, with $a$ the average atomic distance. Going from high to low temperature, the relaxation time diverges: $t_{\alpha}=7$, $243$ and $10999$.

ISs are generated along the MD run by instantaneously quenching configurations at logarithmic increasing time intervals [see Fig.~\ref{fig:method}(a)]. MD configurations are relaxed with balanced forces on each atom using a conjugate gradient algorithm. The kinematics of supercooled liquids can then be greatly simplified and reduced to a sequence of ISs by removing thermal vibrations. Studying the inherent dynamics allows us to focus on the influence of the underlying potential energy landscape~\cite{doliwa_what_2003,vogel_particle_2004,heuer_exploring_2008}. 

In Fig.~\ref{fig:method}(b) and ~\ref{fig:method}(c), we respectively show typical core and long-range thermal rearrangement displacement fields $\overrightarrow{u}_{th}$ between two ISs. We observe a strong localization of $\overrightarrow{u}_{th}$ around rearrangements. The displacement fields in the core of these localized excitations show a wide range of complex behaviors, from single atomic jumps to more spread and intricate patterns. However, a striking feature of the long-range $\overrightarrow{u}_{th}$ is the presence of quadrupolar symmetries~\cite{lemaitre_structural_2014,nishikawa_relaxation_2022} around the rearrangements as shown in Fig.~\ref{fig:method}(c). This is a characteristic feature of Eshelby's plastic shear transformations, more commonly discussed in the context of amorphous plasticity~\cite{Tanguy-EPJE06,maloney_amorphous_2006}. This is a strong hint of the local shear character of thermally activated events in supercooled liquids~\cite{widmer-cooper_central_2009}. We also observe exchange events between atoms, without long-range elastic response, that remain in the minority~\cite{SM}.

\begin{figure*}
   \includegraphics[width=1.95\columnwidth]{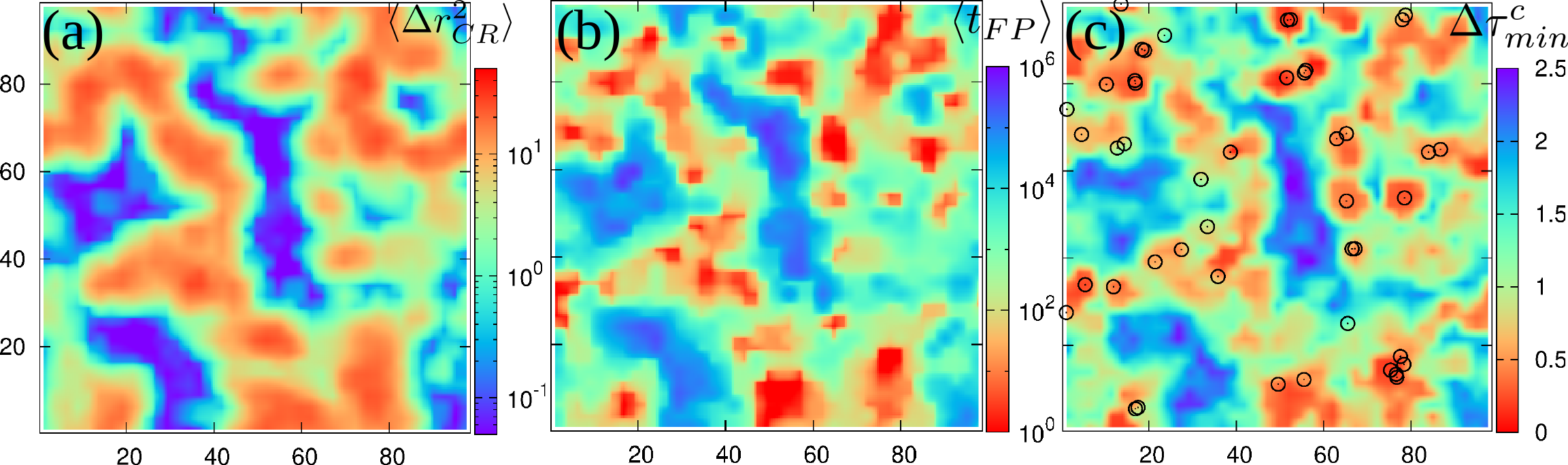}
   \caption{\label{fig:map} Dynamical propensity $\langle\Delta r^{2}_{CR}(t=t_{\alpha})\rangle$ (a), average first passage time $\langle t_{FP}\rangle$ (b) and residual plastic strength in the softest direction $\Delta\tau^c_{min}$     (c) at $T=0.32$. The open circles correspond to the location of the first $50$ isolated thermal rearrangements.}
\end{figure*}

{\em Local shear test method. -- } To study the influence of mechanical properties of the initial structure on the dynamics, we perform LSTs~\cite{barbot_local_2018} in the ISs at time $t=0$. The method consists of shearing a local circular inclusion at position $\protect\ora{r}$ by applying affine pure shear boundary conditions to the surrounding atoms in a shear direction $\theta$ (see Fig.~\ref{fig:method}). The atoms inside are deformed via the quasistatic athermal method~\cite{maloney_subextensive_2004} and can relax nonaffinely. The inclusion is loaded until a drop in local stress indicates the first plastic instability~\cite{SM}. We extract two quantities from the LSTs: the residual plastic strength $\Delta\tau^{c}(\protect\ora{r},\theta)$ and the plastic instability displacement field corresponding to the stress drop $\protect\ora{u}_{LST}(\protect\ora{r},\theta)$. These two observables are sampled with inclusion radii $R_{free}=5$, at positions $\protect\ora{r}$ on a regular square grid of lattice parameter $R_{cut}$ and in $18$ directions $\theta$, every $\Delta\theta=10^{\circ}$.

The values of $\Delta\tau^{c}$ statistically increase when decreasing the parent temperature~\cite{SM}, showing that this observable is able to capture the energy barrier's increase with the decrease of the temperature, i.e. the liquid fragility~\cite{cavagna_supercooled_2009,kallel_evolution_2010}. In the following, we show that $\Delta\tau^{c}$ also determines most of the spatial fluctuations of the dynamics. While isotropy prevails in a statistical sense, the values of the directional $\Delta\tau^{c}(\protect\ora{r},\theta)$ strongly depend on the local shear directions and positions, thus reflecting the disordered character of the amorphous structure~\cite{barbot_local_2018,patinet_origin_2020}. For liquids, the thermally induced excitations are statistically isotropic, so one would expect the onset of local relaxations to be dominated by the residual strengths along the weakest directions: $\Delta\tau^c_{min}(\protect\ora{r})=\min_{\theta}\Delta\tau^{c}(\protect\ora{r},\theta)$. We therefore consider in the following these values in local softest directions $\theta_{min}(\protect\ora{r})$ and compare them with the system dynamics.

{\em Dynamical observables. -- } To emphasize the influence of the structure, we perform MD simulations in the isoconfigurational ensemble~\cite{widmer-cooper_how_2004}. In this ensemble, multiple independent runs are started using the same initial configuration, but the particles’ momenta are drawn from a Maxwell-Boltzmann distribution for each run. For each starting configuration, $100$ replica are simulated. The dynamic propensity $\langle \Delta r^{2}_{CR}(t)\rangle$ is computed as a function of time by averaging the CR square displacement over the isoconfigurational ensemble of each atom~\cite{li_local_2022}. The propensity is finally coarse grained on the same regular grid as $\Delta\tau^c_{min}$ with a coarse-graining length equal to $R_{cut}$ to compare these quantities on an equal footing. A $\langle \Delta r^{2}_{CR}(t)\rangle$ map is reported in Fig.~\ref{fig:map}(a). It shows the characteristic dynamical heterogeneities~\cite{ediger_spatially_2000,keys_excitations_2011}, with some broadening due to the coarse-graining procedure~\cite{SM}.

Although valuable, the information provided by $\langle\Delta r^{2}_{CR}(t)\rangle$ results from an accumulation of rearrangements which cannot be analyzed separately. To address this issue, and detect the positions of rearrangements, we compute the linear force response $\overrightarrow{f}_{lin}$ between two ISs~\cite{lemaitre_structural_2014}. From time $t_{0}$ to $t$, $\overrightarrow{f}_{lin}(t,t_{0})=H_{t_{0}}\cdot\overrightarrow{u}_{th}(t,t_{0})$ where $H_{t_{0}}$ is the Hessian matrix evaluated in the IS at $t_{0}$. We then consider the norm $\rVert\overrightarrow{f}_{lin}\rVert$ on each atom. If the atom motions correspond to the linear response of the system, $\rVert\overrightarrow{f}_{lin}\rVert$ is equal to zero because of mechanical balance. On the other hand, large $\rVert\overrightarrow{f}_{lin}\rVert$ are expected in the highly distorted, nonlinear, regions. This is what is observed in Fig.~\ref{fig:method}(c) where $\rVert\overrightarrow{f}_{lin}\rVert$ features large values at the cores of the thermal rearrangements while vanishing in the rest of the system. Although it requires working in ISs, several advantages of this method should be stressed. First, it is sensitive and $\rVert\overrightarrow{f}_{lin}\rVert$ spans several orders of magnitude between rearranged and elastically disturbed regions. Second, it erases the long-range linear elastic fields induced by rearrangements. Finally, it is not affected by rigid body motions.

\begin{figure}[h]
\includegraphics[width=0.99\columnwidth]{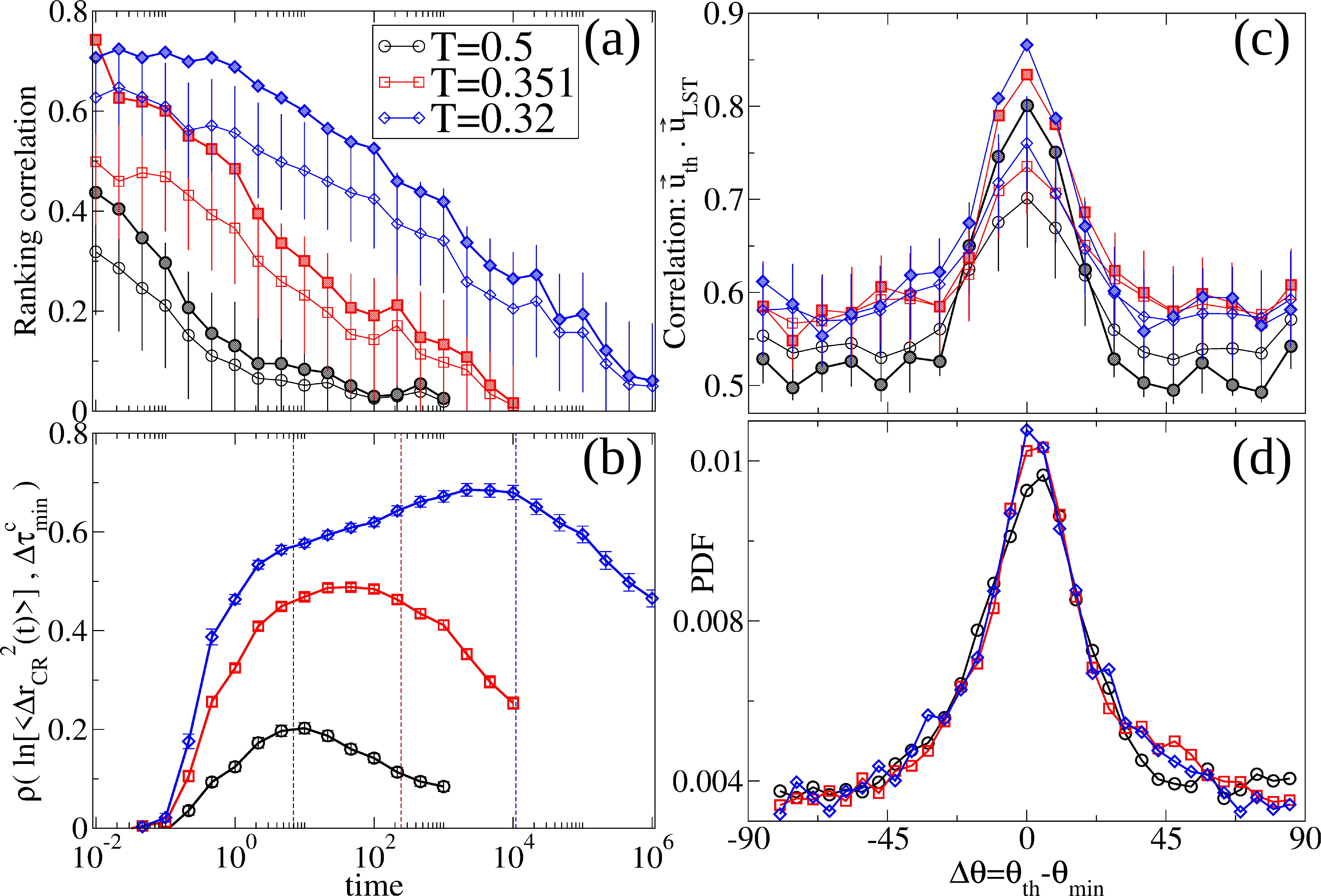}
\caption{\label{fig:correlation} Correlation between the thermal rearrangements and the initial residual plastic strengths in the weakest directions $\Delta\tau^{c}_{min}$. Open and filled symbols correspond to the average and the median, respectively. Ranking (a) and Pearson correlations (b) as a function of time. (c) Correlation between thermal and LST displacement fields [see Fig.~\ref{fig:method}(b)] as a function of the angular difference $\Delta\theta=\theta_{th}-\theta_{min}$ between the shear strain of the thermal rearrangement and the softest direction. (d) $\Delta\theta$ probability distribution function.}
\end{figure}

We choose to consider as being rearranged the atoms for which $\rVert\overrightarrow{f}_{lin}(t,t_{0})\rVert>f^{*}$, with $f^{*}=7$~\cite{SM}. This allows us to define the local first passage time $t_{FP}$ when this criterion is fulfilled for the first time during a trajectory starting from $t_{0}=0$. For each time lag $\Delta t$ and for each isolated clusters of atoms with $\rVert\overrightarrow{f}_{lin}(\Delta t=t-t_{0})\rVert>f^{*}$, the locations of the successive isolated thermal rearrangements are extracted from the positions of the atom carrying the largest $\rVert\overrightarrow{f}_{lin}\rVert$. For each event, we also record the local maximum shear strain direction denoted $\theta_{th}$ from the local strain field~\cite{barbot_shearband_2020}. The number of events between two ISs increases with temperature and time window $\Delta t$. The isoconfigurational averages of first passage times $\langle t_{FP}\rangle$ reported in Fig.~\ref{fig:map}(b) show again characteristic dynamical heterogeneities of supercooled liquids. The comparison of $\langle\Delta r^{2}_{CR}(t=t_{\alpha})\rangle$ and $\langle t_{FP}\rangle$ shows striking similarities, meaning that parts of the short and long dynamics are correlated. In our view, however, $\langle t_{FP}\rangle$ provides an easier grasp of the initial local structural relaxation timescale compared with accumulated mean square displacements.

{\em Correlation between dynamics and $\Delta\tau^c_{min}$. -- } In Fig.~\ref{fig:map}, we compare the $\langle\Delta r^{2}_{CR}(t=t_{\alpha})\rangle$, $\langle t_{FP}\rangle$ and $\Delta\tau^{c}_{\min}$ maps. Their similarity is remarkable. The softer the region, the faster its relaxation. As an example, the first $50$ isolated thermal events of a particular MD trajectory are reported in Fig~\ref{fig:map}c. They are systematically located in the softest areas. More quantitatively, to analyze the correlation between the rearrangement locations and the initial structure, we report in Fig.~\ref{fig:correlation}(a) the ranking correlation defined in Ref.~\cite{patinet_connecting_2016} from the cumulative distribution function $F_{X}$ as $C(t)=1-2 F_{X}(X,t)$ where $X=\Delta\tau^{c}_{\min}$ is the value of the field at $t=0$ where the thermal event takes place at $t$.  $C(t)$ shows that the correlation is larger for the smallest temperatures and decreases over time, the system losing the memory of its original structure.

The Pearson correlation $\rho$ between the fields $\Delta\tau^{c}_{\min}(t=0)$ and $\ln[\langle\Delta r^{2}_{CR}(t)\rangle]$ is plotted as a function of time in Fig.~\ref{fig:correlation}(b). We observe that the correlation increases with time as the successive thermal relaxations accumulate, with a  peak around $t_{\alpha}$. The correlation with the $\langle t_{FP}\rangle$ field is quantitatively close to that with $\langle\Delta
r^{2}_{CR}(t=t_{\alpha})\rangle$~\cite{SM}. Remarkably, for the lowest temperature, we find a maximum correlation $\rho\approx 0.7$ comparable to the maximum correlations found in literature such those obtained from machine learning~\cite{bapst_unveiling_2020,boattini_autonomously_2020} and to the order parameter proposed in Ref.~\cite{tong_structural_2019} (see Supplemental Material for a detailed comparison~\cite{SM}). 

\begin{figure}
\includegraphics[width=0.95\columnwidth]{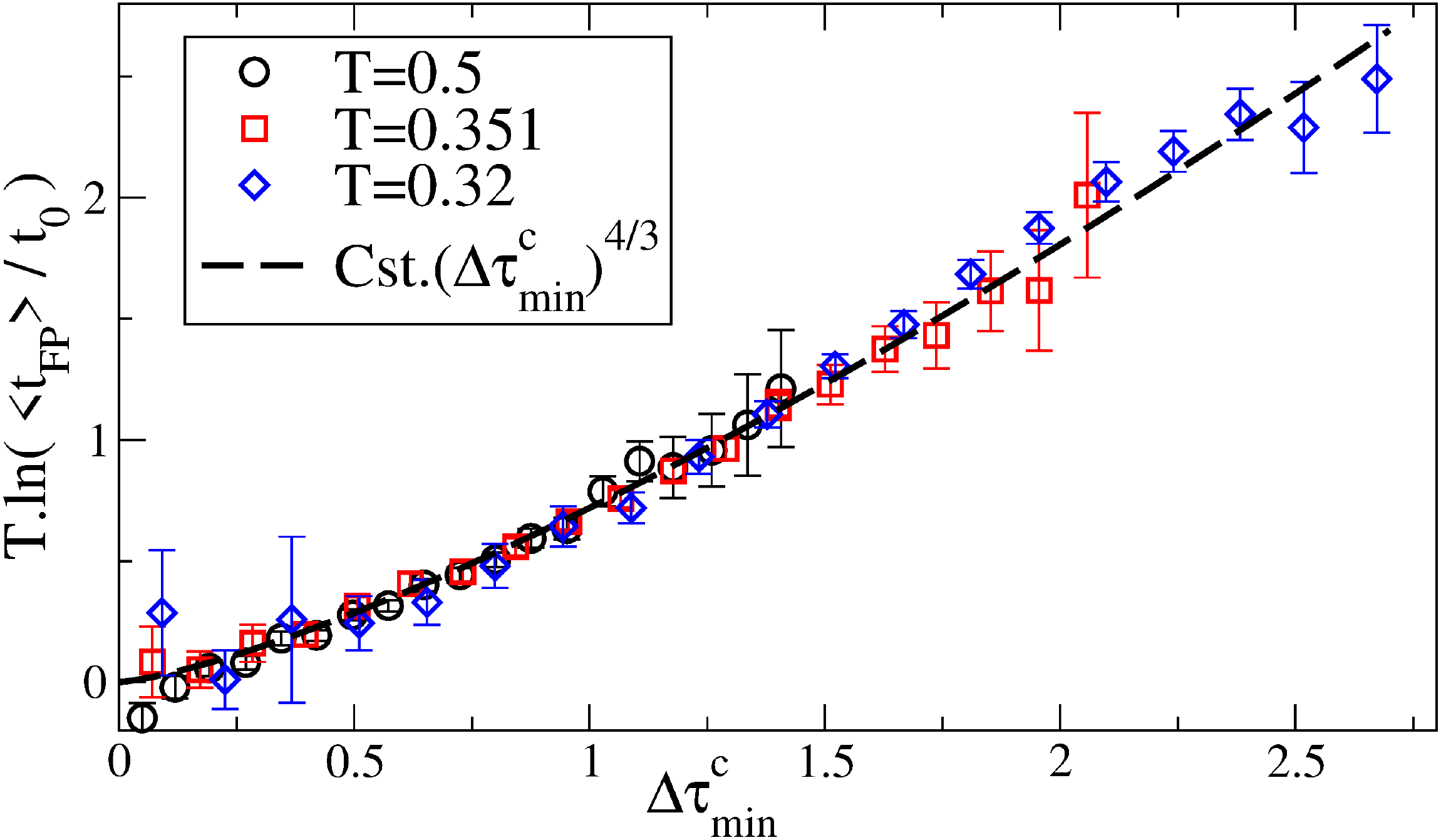}
\caption{\label{fig:model}
Energy barrier $\Delta U$ estimated from the average first passage time $\langle t_{FP}\rangle$ as a function of the residual plastic strength $\Delta\tau^c_{min}$. $t_0$ is a tunable parameter fitted to collapse the data. The dashed line shows the scaling law found in Ref.~\cite{kapteijns_nonlinear_2020}.}
\end{figure}

We go one step further and directly evaluate the correlation between
$\overrightarrow{u}_{th}$ and $\overrightarrow{u}_{LST}$ in the core regions of the first passage rearrangements. 
An example of such a comparison reported in Fig.~\ref{fig:method}(b) shows a striking agreement
between the two fields. The correlation is calculated as the maximum
over all LST directions $\theta$ of the dot product of the normalized vector fields $\overrightarrow{u}_{th}$ and
$\overrightarrow{u}_{LST}$. Figure \ref{fig:correlation}(c) shows correlation as a function of the orientation deviation $\Delta\theta=\theta_{th}-\theta_{min}$. We observe a maximum when the thermal rearrangements are
aligned with the softest directions, i.e. for $\Delta\theta=0$. This maximum of correlation obtained on the full set of first passage thermal rearrangements reaches significantly high values with a median in the range $[0.8-0.86]$ while the mean lies a bit below in the range $[0.7-0.76]$. This slight contrast reflects the fact that the distribution presents a peak close to 1, i.e. to quasi-identity between plastic and thermal events. 

In addition to barrier heights, this result shows that the mechanisms of shear transformation and thermal relaxation share remarkable similarities. Besides the correlation, we report in Fig.~\ref{fig:correlation}(d) the $\Delta\theta$ probability distribution showing a clear peak around $\Delta\theta=0$, which means that thermal rearrangement's shear directions are more frequent in the vicinity of softest directions. This last observation is fully consistent with the interpretation of fast dynamics in the soft regions of ISs, whose hop frequencies are dominated by the weakest directions. 

{\em Scaling of energy barriers. -- } Once the correlation is established, we discuss now the quantitative relation between $\langle t_{FP}\rangle$ and $\Delta\tau^c_{min}$. The effective activation energy is fitted assuming an Arrhenius law in the form $\langle t_{FP}\rangle=t_{0}e^{\Delta U/T}$ with $t_{0}$ a prefactor and $\Delta U$ the energy barrier. The actual temperature dependence of the characteristic time $t_0$ is disregarded, and $t_0$ is used as a tunable parameter allowing one to collapse the results obtained at different temperatures. As shown in Fig.~\ref{fig:model}, we find a scaling of barriers as a function of the residual plastic strengths $\Delta U \sim (\Delta\tau^{c}_{min})^{\beta}$. If an exponent $\beta=3/2$, compatible with the catastrophe theory~\cite{johnson_universal_2005,maloney_energy_2006,Egami-NatComm14}, can not be totally ruled out, the fit of our data gives $\beta=1.33\pm 0.03$. Interestingly, this result is in quantitative agreement with $\beta=4/3$, a scaling that can be deduced from the relations $\Delta U\sim\omega^{4}$ and $\Delta\tau^{c}\sim\omega^{3}$ found in Ref.~\cite{kapteijns_nonlinear_2020} [see Eqs. (35) and (37) therein] where $\omega$ is the frequency of the nonlinear quasilocalized excitations in quiescent glasses. It is to our knowledge the first time it has been obtained locally at finite temperature. This scaling relation can be understood as a reflection of the sensitivity of the lowest barriers to stress, which is identified here in the local softest directions by the LST.

{\em Concluding remarks and perspectives. -- } In this Letter, we have established a strong correlation between residual plastic strengths in the softest direction of the initial IS and the dynamics of glass-forming liquids. The softer the zones of ISs, the faster the local region of liquids. An advantage of this picture over purely geometrical approaches is that it provides a simple real-space interpretation of relaxations as localized shear events in soft directions. Among the possible transition paths, a local shear allows particles to escape from their cage for moderate long-range elastic energy cost. As a local stress, $\Delta\tau^{c}_{min}$ is a naturally coarse-grained quantity~\cite{berthier_structure_2007,tong_revealing_2018}.
Shear transformations thus involve cooperative motions in the spirit of an Adam-Gibbs-style scenario~\cite{cavagna_supercooled_2009,berthier_self-induced_2021} and long-range generated elastic interactions providing intuitive support to the dynamic facilitation picture~\cite{chacko_elastoplasticity_2021,hasyim_theory_2021,scalliet_excess_2021}.  By pointing to the central role played by the mechanics of ISs at low temperatures, these results are useful for understanding the onset of rigidity and the theoretical description of the glass transition.

\begin{acknowledgments}
The present work was initiated during the PhD thesis of the late Matthias Lerbinger. A.B., S.P. and D.V. dedicate this article to his memory.
\end{acknowledgments}

\bibliography{bilbio_LBVP_2022}

\bigskip
\bigskip

\onecolumngrid
\begin{center}
\textbf{\large Supplemental Material to: ``Relevance of shear transformations in the relaxation of supercooled liquids''}
\end{center}
\twocolumngrid

\setcounter{equation}{0}
\setcounter{figure}{0}
\setcounter{page}{1}

\renewcommand{\bibnumfmt}[1]{[SM#1]}
\renewcommand{\theequation}{SM\arabic{equation}}
\renewcommand{\thefigure}{SM\arabic{figure}}

\section*{System}

The system's density is constant and equals $10^4/(98.8045)^2 \approx 1.02$. Periodic boundary conditions are imposed on square boxes of linear dimensions $L=98.8045$. Following~\cite{barbot_local_2018}, we choose the composition so that the ratio of the number of large (L) and small (S) particles, of equal mass, is equal to $N_{L}/N_{S}=(1+\sqrt{ 5})/4$. The two types of atoms interact via standard Lennard-Jones $6-12$ interatomic potentials employed in~\cite{falk_dynamics_1998} with zero-energy interatomic distances $\sigma_{LL}=2\sin(\pi/5)$, $\sigma_{SS}=2\sin(\pi/10)$ and $\sigma_{LS}=1$, and bond strengths $\epsilon_{LL}=\epsilon_{SS}=1/2$ and $\epsilon_{LS}=1$.

In our work, these potentials have been slightly modified to be twice continuously differentiable functions. The expression of the Lennard-Jones potential for interatomic distances greater than $R_{in}=2\sigma$ is replaced by a quartic function which vanishes at a cutoff distance $R_{cut}=2.5\sigma$. For two atoms $i$ and $j$ separated by a distance $r_{ij}$, the potential is written:

\begin{equation}
\label{eq:interatomicpotential}
U(r_{ij}) =
\begin{cases}
4\epsilon_{\alpha\beta}\left[ \left(\frac{\sigma_{\alpha\beta}}{r_{ij}}\right)^{12}-\left(\frac{\sigma_{\alpha\beta}}{r_{ij}}\right)^{6}\right]+A, \quad \text{  for } r_{ij}<R_{in}\\
\sum_{k=0}^{4}C_{k}(r_{ij}-R_{in})^{k}, \quad \text{for } R_{in}<r_{ij}<R_{cut}\\
0, \qquad \qquad \qquad \qquad \qquad \qquad \ \ \text{for } r_{ij}>R_{cut},
\end{cases}
\end{equation}
with
\begin{subequations}
\label{eq:interatomicpotentialcoeff}
\begin{align*}
A&=C_{0}-4\epsilon_{\alpha\beta}\left[\left(\frac{\sigma_{\alpha\beta}}{R_{in}}\right)^{12}-\left(\frac{\sigma_{\alpha\beta}}{R_{in}}\right)^{6}\right]\\
C_{0}&=-(R_{cut}-R_{in})(3C_{1}+C_{2}(R_{cut}-R_{in}))/6 \\
C_{1}&=24\epsilon_{\alpha\beta}\sigma_{\alpha\beta}^6(R_{in}^6-2\sigma_{\alpha\beta}^6)/R_{in}^{13} \\
C_{2}&=12\epsilon_{\alpha\beta}\sigma_{\alpha\beta}^6(26\sigma_{\alpha\beta}^6-7R_{in}^6)/R_{in}^{14} \\
C_{3}&=-(3C_{1}+4C_{2}(R_{cut}-R_{in}))/(3(R_{cut}-R_{in})^2) \\
C_{4}&=(C_{1}+C_{2}(R_{cut}-R_{in}))/(2(R_{cut}-R_{in})^3), \\
\end{align*}
\end{subequations}
where the subscripts $\alpha,\beta$ denote the interacting species, S or L.

\section*{Local shear test}

\begin{figure}
\includegraphics[width=0.95\columnwidth]{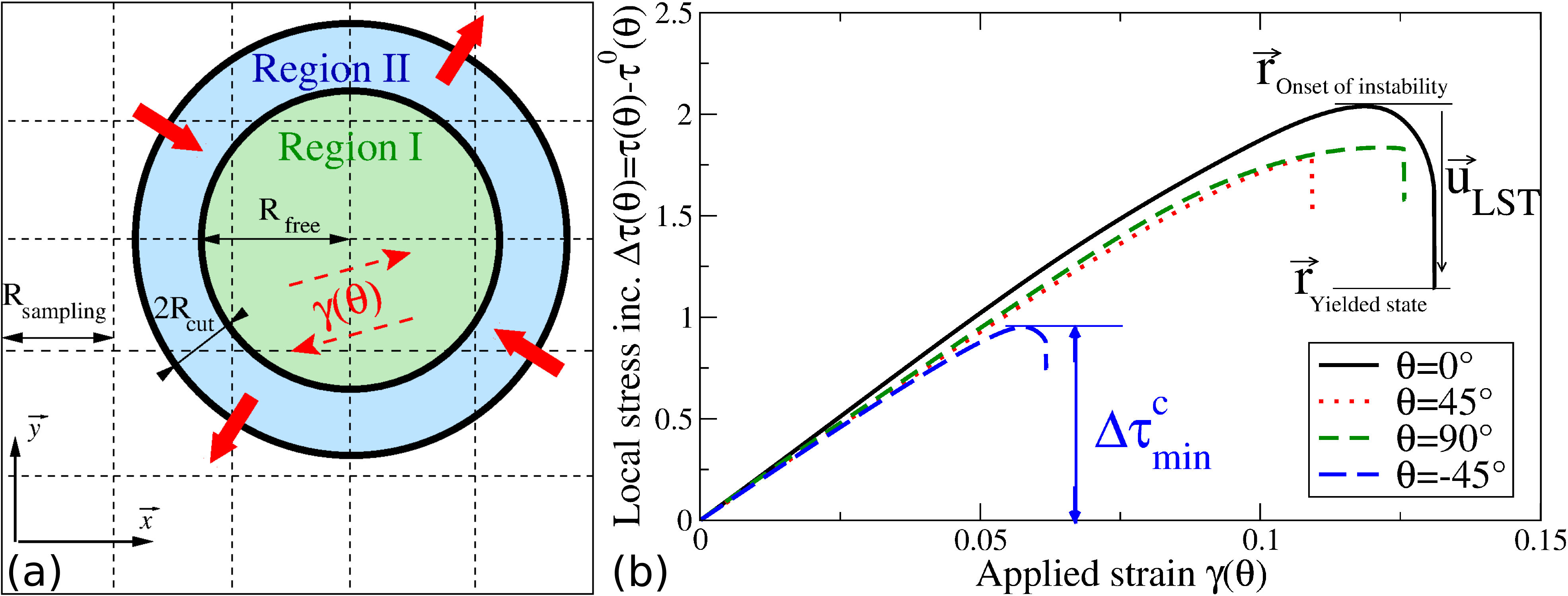}
\caption{\label{fig:SM_LTS} (a) Schematic drawing of the local shear test on a regular square grid of mesh size $R_{sampling}$. Region I (of radius $R_{free}$) is fully relaxed, while region II (of width $2R_{cut}$) is forced to deform following an affine pure shear deformation in the loading direction $\theta$. (b) Typical local stress increment-strain curves of Region I for $\theta=0$, $45$, $90$, and $-45^{\circ}$. The residual plastic strengths in the softest direction $\Delta\tau^c_{min}$ is found for $\theta_{min}=-45^{\circ}$. The two states used to compute the displacement $\protect\ora{u}_{LST}$ are illustrated for $\theta_{min}=0^{\circ}$.}
\end{figure}

Residual plastic strengths $\Delta\tau^{c}$ are computed in the initial inherent states (ISs) by following the local shear test (LST) method developed in~\cite{barbot_local_2018}. The principle of the method is illustrated in the Fig.~\ref{fig:SM_LTS}(a). It consists in forcing the atoms outside a circular inclusion [Region II in Fig.~\ref{fig:SM_LTS}(a)] of size $R_{free}$ to deform following an affine pure shear in the direction $\theta$ up to the first plastic instability. It thus allows to locally probe the nonlinear mechanical response. The inclusion is loaded with an athermal quasi-static protocol~\cite{maloney_amorphous_2006} which consists in incrementally applying small affine deformation step $\Delta\gamma=10^{-4}$ followed by a relaxation step, performed via the conjugate gradient method. Only the atoms of the inclusion [Region I in Fig.~\ref{fig:SM_LTS}(a)] are relaxed and can deform in a nonaffine way. Plastic rearrangements are therefore forced to occur in the inclusion, and the local yield stress $\tau^{c}(\theta)$ can be identified.

The local plasticity criterion, occurring at critical stress $\tau^{c}(\theta)$, corresponds to the shear stress drop in the loading direction $\theta$. Even at rest, glasses feature non-zero internal stresses due to frustration in the amorphous solids. A more relevant quantity for linking local properties to plastic activity is
thus the amount of stress needed to trigger plastic rearrangement. The initial local shear stress state in the glass $\tau^0(\theta)$ is therefore subtracted from the critical stress to obtain the increase in local stress, which triggers an instability. The residual plastic strength thus writes $\Delta\tau^c(\theta)=\tau^{c}(\theta)-\tau^{0}(\theta)$

To get the full spatial field $\tau^{c}(\theta)$, we repeat the LST on a regular square grid with lattice parameter $R_{Sampling}=R_{cut}$. Non-interacting atoms, i.e. located at a distance larger than $(R_{free}+2R_{cut})$ from the centre of the probed inclusion, are removed during the local shear loading simulations, thus accelerating the calculation. In this study, we chose $R_{free}=5$. It was shown in~\cite{barbot_local_2018} that this length optimizes the prediction power of the locations of plastic rearrangements under mechanical loading in athermal conditions. Interestingly, this inclusion size is close to the optimal coarse-graining length found in~\cite{tong_structural_2019,tong_role_2020} to link structure and dynamics at the lowest temperatures of supercooled liquids. 

At this scale, the amorphous system is heterogeneous, and the residual plastic strengths are not the same for all the imposed shear orientations $\theta$ as reported in Fig.~\ref{fig:SM_LTS}(b). We sample the mechanical response using pure shear loading conditions along $18$ different directions uniformly distributed between $\theta=0$ and $\theta=170^{\circ}$. The displacement field corresponding to the local plastic instability $\overrightarrow{u}_{LST}$ are also computed as the difference between the positions of the atoms in the inclusions at the onset of the instability and after the stress drops.

\section*{Statistics of residual plastic strengths in the inherent states}

\begin{figure}
\includegraphics[width=0.95\columnwidth]{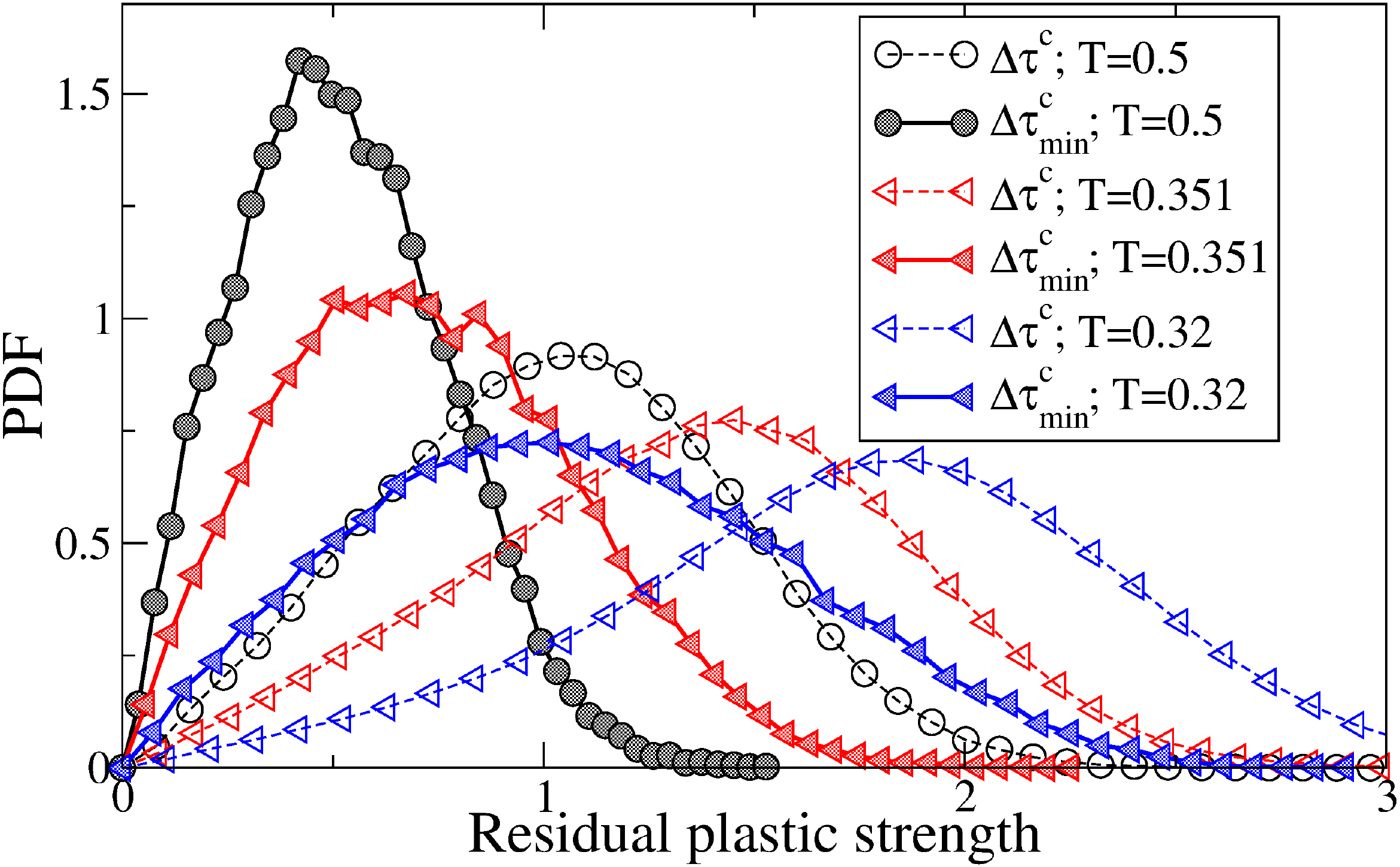}
\caption{\label{fig:SM_residual_plastic_strength_PDF} Probability distribution function of residual plastic strengths in all directions $\Delta\tau^c(\theta)$ (dashed lines) and of their minima over orientation $\Delta\tau^c_{min}$, i.e. in the weakest direction, ${\theta_{min}}$ (solid lines).}
\end{figure}

We report in the Fig.~\ref{fig:SM_residual_plastic_strength_PDF} the $\Delta\tau^{c}(\theta)$ distributions obtained in the initial ISs for the three simulated parent temperatures. The local slip thresholds increase as temperature is decreased, the glasses becoming harder and stiffer~\cite{barbot_local_2018}. This hardening effect stems from increasing energy barriers as systems sink lower into the potential energy landscape. Fig.~\ref{fig:SM_residual_plastic_strength_PDF} also shows the distributions of the threshold minima over all the sampled directions $\theta$ for each site, $\Delta\tau^{c}_{min}=\min_{\theta}\Delta\tau^{c}(\theta)$. This filter over orientations shifts the distributions downwards and the ranking between the different temperatures remains unchanged.

\section*{Effect of propensity coarse-graining}

\begin{figure}
\includegraphics[width=0.95\columnwidth]{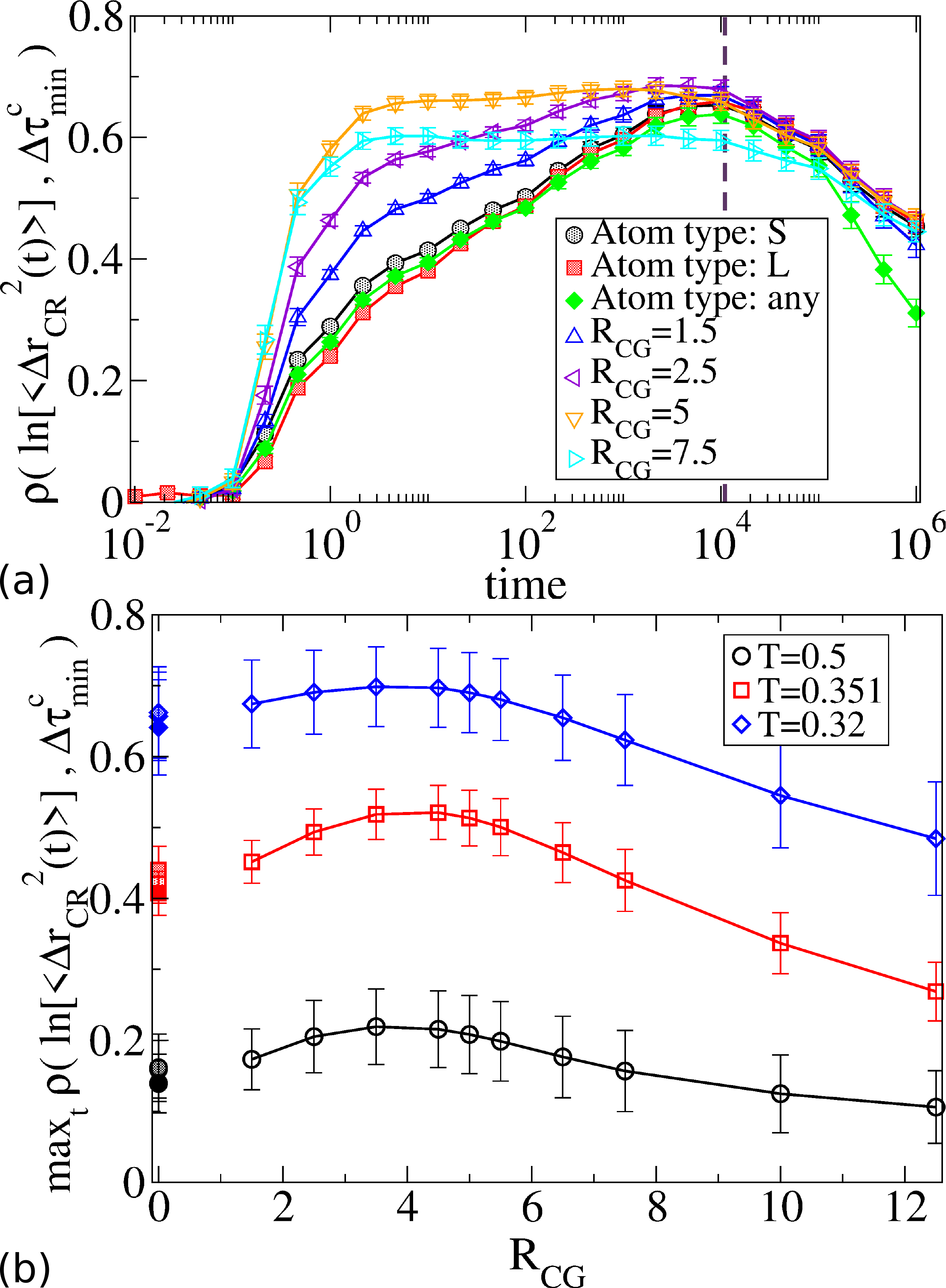}
\caption{\label{fig:SM_correlation_vs_RCG} Pearson correlation $\rho$ between $ln(\langle\Delta r^{2}_{CR}\rangle)$ and $\Delta\tau^c_{min}$. The open and filled symbols correspond to the coarse-grained and single atom propensity, respectively. (a) $\rho$ as a function of time for $T=0.32$ for different coarse-graining scales $R_{CG}$. (b) maximum of $\rho$ over time as a function of $R_{CG}$ for the three temperatures.}
\end{figure}

The effect of propensity coarse-graining is analyzed by calculating the correlation between $\Delta\tau^c_{min}$ and $\langle\Delta r^{2}_{CR}\rangle$ for different scales of coarse-graining length $R_{CG}$. Since the propensity depends on the atom type, we compute the correlation for both atom types when the propensity is not coarse-grained, i.e., with $R_{CG}=0$. In this situation, we consider the atom (of type L, S or both) closest to the grid points on which the residual plastic strengths are sampled. The results are reported in Fig.~\ref{fig:SM_correlation_vs_RCG} as a function of time and $R_{CG}$.

The results discussed in the main text are qualitatively similar for different $R_{CG}$: the correlation increases to reach a maximum located around $t_{\alpha}$. Closer inspection, however, shows slight variations. The correlation grows faster with a larger $R_{CG}$. On the other hand, beyond a scale $R_{CG}=5$, the maximum correlation starts to decrease, the analysis losing the spatial resolution of the mobility. As expected, the maximum correlations obtained are slightly higher, with a subsequent slower decrease, for isolated atoms when considering the chemical species (type $S$ and $L$) than without taking it into account (any). The maximum correlation optimum is found around $R_{CG}\approx 4$ and shows a weak temperature dependence. Finally, we note that the $R_{CG}$ employed in the main text has not been optimized to maximize the correlation but only to compare an indicator of the dynamics, here the propensity, with $\Delta\tau^c_{min }$ using similar spatial sampling.

\section*{Comparison with the maximum correlations found in literature}

In order to quantify the performance of our analysis, we compare in Tab.~\ref{tab:maximum_correlation} the best correlations obtained in recent numerical references. We first note some methodological differences from the present work that prevent a truly quantitative comparison. A Spearman's rank correlation coefficient is employed in~\cite{tong_structural_2019,boattini_autonomously_2020}. A three-dimensional system is considered in~\cite{bapst_unveiling_2020,boattini_autonomously_2020} and the analysis focuses on one of the two types of particles of their binary systems. Instead of the propensity, dynamics is quantified by the microscopic (local) relaxation time $t_{\alpha}^{i}$ for particle $i$ in~\cite{tong_structural_2019}. Finally, we exclude from the comparisons the short vibration times for which an almost perfect correlation was found in~\cite{bapst_unveiling_2020}, the machine learning method having learned the ballistic regime. We observe a comparable maximum correlation between dynamics and structure despite these differences.

\begin{table*}
\caption{\label{tab:maximum_correlation} Maximum correlations $\rho$ between initial structure and dynamics for different numerical references.}
\begin{ruledtabular}
\begin{tabular}{c|c|c|c|c|c}
Reference & System & Structural indicator/Method & Dynamical observable & Correlation definition & $\rho$ value \\
\hline
\hline
\cite{boattini_autonomously_2020} & 3D binary hard spheres  & unsupervised machine learning  & propensity around $t_\alpha$ & Spearman & 0.74 \\
& & artificial neural network & & & \\
\hline
\cite{bapst_unveiling_2020}       & 3D binary Lennard–Jones & supervised machine learning    & propensity around $t_\alpha$ & Pearson  & 0.64 \\
								  & 						& graph neural networks 		 & 			  & 		& \\
\hline
\cite{tong_structural_2019}       & 2D binary Lennard–Jones & $\Theta_{CG}$ coarse-grained order   		 & microscopic $t_{\alpha}^{i}$ & Spearman        & $0.78\pm0.01$ \\
								  & (same as this work)	    & parameter computed in ISs 		 & 			  & 		& \\
\hline
This work                         & 2D binary Lennard–Jones & $\Delta\tau^c_{min}$           &  $\langle t_{FP}\rangle$ & Pearson & $0.71\pm0.05$\\
\end{tabular}
\end{ruledtabular}
\end{table*}

In order to establish a comparison as quantitative as possible, we have also rigorously reproduced the method proposed in Ref.~\cite{tong_structural_2019} in our system. We calculate the local relaxation times $t_{\alpha}^{i}$ and compare them to the proposed steric bond order $\Theta_{CG}$. Note that this method implies a coarse-graining length $R_{CG}$ chosen to maximize the correlation. In agreement with~\cite{tong_structural_2019}, we observe a high Spearman correlation, with an optimal coarse-graining length which increases with decreasing temperature. For the lowest temperature, at $T=0.32$, we find an optimal length $R_{CG}=4.5$, again very close to the inclusion size $R_{free}$ employed in our LST method. We note, however, that, contrary to the results reported in ~\cite{tong_structural_2019}, the correlation is significantly better when $\Theta_{CG}$ is calculated in the ISs rather than in the thermalized states (giving $\rho=0.6\pm0.01$ in that case).

\section*{Thermal rearrangement detection}

\begin{figure}
\includegraphics[width=0.95\columnwidth]{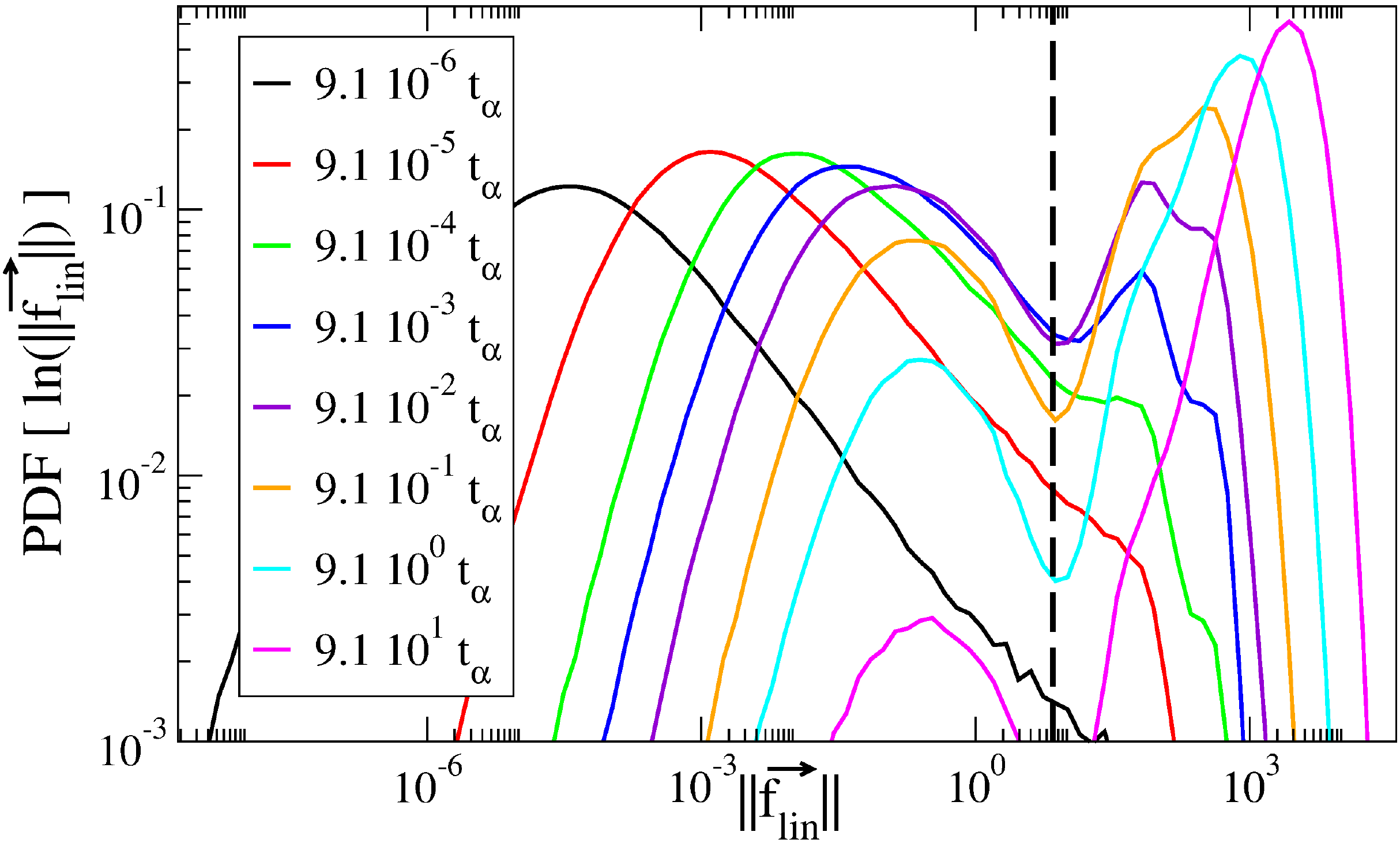}
\caption{\label{fig:SM_detection} Probability distribution functions of the linear force norms $\rVert\protect\ora{f}_{lin}\rVert$ computed in the ISs for different lag times at $T=0.32$. The vertical dashed line is the detection threshold $f^{*}=7$ used in this work.}
\end{figure}

In support of our thermal rearrangement detection method, we present the distribution functions of the linear response computed through the linear force norms $\rVert\overrightarrow{f}_{lin}\rVert$ described in the main text. The force norm $\rVert\overrightarrow{f}_{lin}\rVert$ varies by several orders of magnitude in space and time. We report in Fig.~\ref{fig:SM_detection} the distributions of $\ln(\rVert\overrightarrow{f}_{lin}\rVert)$ in order to apprehend the scales of variation of the linear response~\cite{lemaitre_plastic_2007,barbot_shearband_2020}. At short times, the distribution functions exhibit a peak for infinitesimal values of $\rVert\overrightarrow{f}_{lin}\rVert$ corresponding to regions not yet rearranged. On the other hand, completely rearranged systems show at long times a peak at large values of $\rVert\overrightarrow{f}_{lin}\rVert$. The intermediate distributions shift to the right with increasing lag time between ISs and show a two-peak shape with a crossover located around a value $\rVert\overrightarrow{f}_{lin}\rVert \approx 7$. 

Interestingly, the time at which the two peaks have equal heights, interpreted as a sign of a phase transition from an initial state to a rearranged state, is found around the relaxation time $t_{\alpha}$. Furthermore, we observe that the crossover position is almost constant over time. We therefore choose a threshold $f^{*}=7$ in order to detect the thermal rearrangements. We verified that the variation of this threshold does not qualitatively change the results presented in this study. We also found that this threshold is the same for all temperatures, although with a slightly less marked minimum in the distributions at higher temperatures.

To understand the origin of this threshold, we rely again on the LST method performed in the soft directions. We compute the maximum of $\rVert\overrightarrow{f}_{lin}\rVert$ over the atoms in the locally probed inclusions from the displacement fields between the initial ISs at rest, and the final yielded states, just after the plastic instabilities. Their probability distribution functions shows a peak just after the threshold $f^{*}$ found in Fig.~\ref{fig:SM_detection}. We, therefore, interpret this value as a characteristic scale corresponding to local shear rearrangements which is fully consistent with our interpretation of relaxations.

\section*{Exchange events}

\begin{figure}
   \includegraphics[width=0.95\columnwidth]{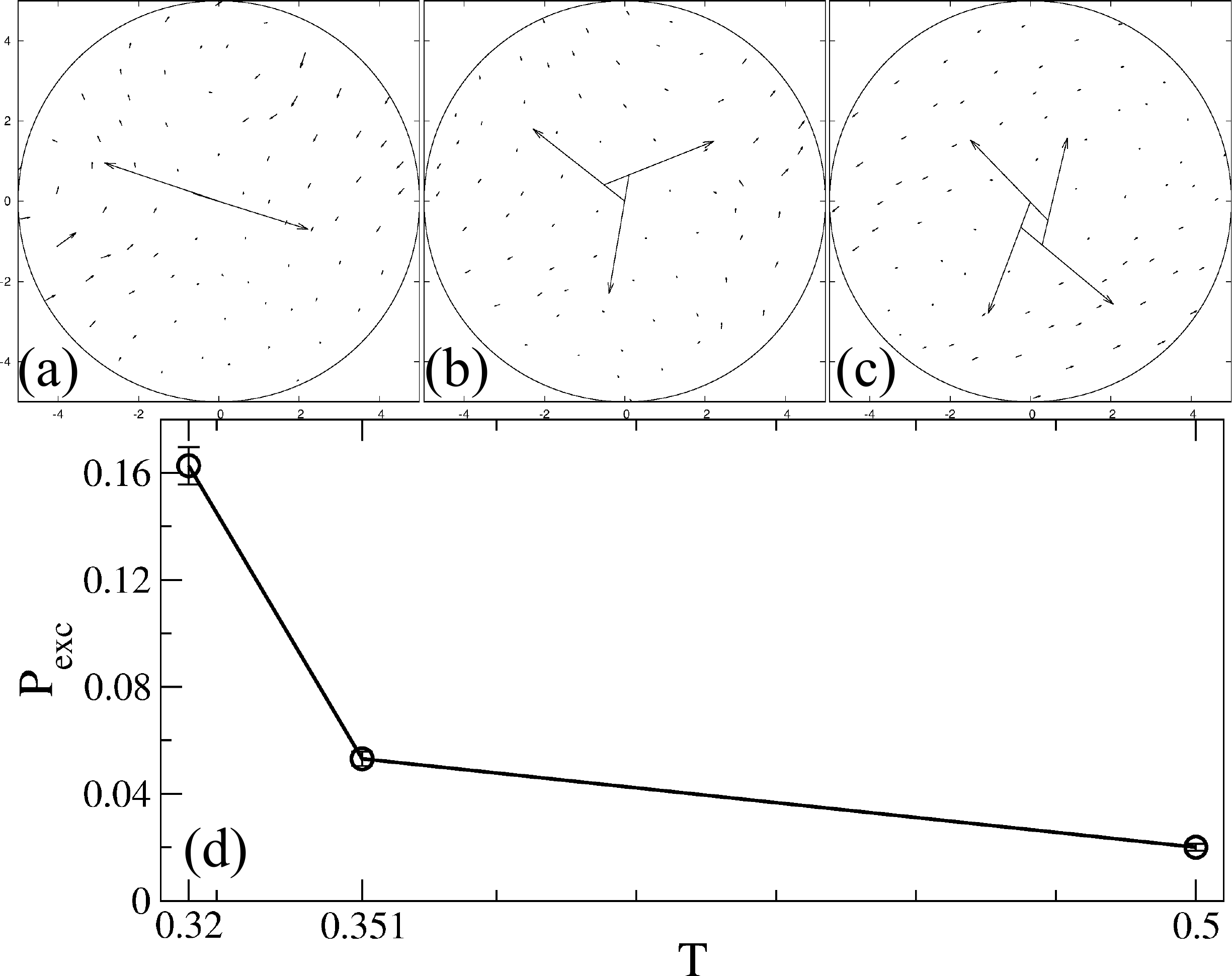}
\caption{\label{fig:SM_exchange_event} Exchange event displacement fields corresponding to 2 (a), 3 (b) and 4 (c) atom processes. (d) Exchange event proportion $P_{exch}$ in the isolated thermal rearrangement population as a function of temperature.}
\end{figure}

Some isolated thermal rearrangements do not fall into the local shear category. These are atom-exchange events that do not create far-field displacements giving rise to a redistribution of stresses. Examples of these types of rearrangement are shown in Fig.~\ref{fig:SM_exchange_event}(a), ~\ref{fig:SM_exchange_event}(b) and ~\ref{fig:SM_exchange_event}(c). Because of the atom exchange, the energy landscape is nevertheless disturbed in a strongly non-linear way, and our detection algorithm can detect these rearrangements unambiguously. 

To differentiate them within the isolated thermal rearrangements population, we observe that the largest displacement are carried by the central atoms participating in the exchange processes. The number of these central atoms is first evaluated from the nearest integer of the local displacement field participation number. A rearrangement is then classified as an exchange event when the sum of displacements over these central jumping atoms is small (i.e. when their displacements nearly form a closed-loop) compared to the largest local displacement. The exchange events are removed from the analysis presented in Fig.~3(c) and 3(d) of the main text.

We find that the proportion of these exchange events $P_{exch}$ among the isolated events defined in the main text increases as the temperature decreases, as shown in Fig.~\ref{fig:SM_exchange_event}. However, this proportion remains small regardless of the temperature. Furthermore, the trend observed in Fig.~\ref{fig:SM_exchange_event} may be questioned because the structure is strongly disturbed at high temperatures, even for short time windows. Therefore, it is possible in this case that events of several types overlap, preventing these exchange events from being identified.

\end{document}